\documentclass[superscriptaddress,preprint,nofootinbib]{article}
\pdfoutput=1 
\usepackage{jheppub}
\usepackage{graphicx}
\usepackage{amssymb}
\usepackage{amsmath}
\usepackage{epstopdf}
\usepackage{subfigure}
\usepackage{color}
\usepackage{mathtools}

\graphicspath{{./Artwork/}}

\def\wb{\bar{w}}
\def\Xb{\bar{X}}
\def\dalpha{{\dot{\alpha}}}
\def\dbeta{{\dot{\beta}}}
\def\tf{{\tilde{f}}}
\def\tlambda{{\tilde{\lambda}}}
\def\hv{{\hat{V}}}
\def\Co{\mathbb{C}}
\def\be{\begin{equation}}
\def\ee{\end{equation}}

\begin{document}
\begin{flushright}
Edinburgh 2013/29
\vspace*{-25pt}
\end{flushright}

\title{The Kinematic Algebras from the Scattering Equations}

\author{Ricardo Monteiro${}^{a}$,}
\author{Donal O'Connell${}^{b}$}
\affiliation{${}^a$Mathematical Institute, University of Oxford, Woodstock Road, Oxford OX2 6GG, UK}
\affiliation{${}^b$Higgs Centre for Theoretical Physics, School of Physics and Astronomy, The University of Edinburgh, Edinburgh EH9 3JZ, Scotland, UK}

\abstract{We study kinematic algebras associated to the recently proposed scattering equations, which arise in the description of the scattering of massless particles. In particular, we describe the role that these algebras play in the BCJ duality between colour and kinematics in gauge theory, and its relation to gravity. We find that the scattering equations are a consistency condition for a self-dual-type vertex which is associated to each solution of those equations. We also identify an extension of the anti-self-dual vertex, such that the two vertices are not conjugate in general. Both vertices correspond to the structure constants of Lie algebras. We give a prescription for the use of the generators of these Lie algebras in trivalent graphs that leads to a natural set of BCJ numerators. In particular, we write BCJ numerators for each contribution to the amplitude associated to a solution of the scattering equations. This leads to a decomposition of the determinant of a certain kinematic matrix, which appears naturally in the amplitudes, in terms of trivalent graphs. We also present the kinematic analogues of colour traces, according to these algebras, and the associated decomposition of that determinant. 
}

\maketitle
\tableofcontents

\section{Introduction}
\label{sec:intro}

Cachazo, He and Yuan \cite{Cachazo:2013gna,Cachazo:2013hca} have recently proposed remarkably compact formulas for the tree-level scattering amplitudes of gluons and gravitons, valid in any number of spacetime dimensions. These are expressed in terms of an integral over points on a sphere, which localises to a sum over $(n-3)!$ contributions, for $n$-particle scattering. The localisation of the integral is associated to the $(n-3)!$ solutions of the so-called scattering equations. This will be reviewed in detail in the next section.

In the case of gauge theory, the new formula can be seen as an extension of the four-dimensional twistor string theory connected prescription developed by Roiban, Spradlin and Volovich \cite{Roiban:2004yf}, following Witten's seminal work \cite{Witten:2003nn}. For gravity, the four-dimensional analogue are the expressions inspired by twistor string theory found recently by Cachazo and Geyer \cite{Cachazo:2012da} and by Cachazo and Skinner \cite{Cachazo:2012kg}; see also \cite{Adamo:2012xe,Feng:2012sy,Bullimore:2012cn,He:2012er,Cachazo:2012pz,Skinner:2013xp,Cachazo:2013zc,Adamo:2013tja,Adamo:2013tca}. The latter developments were partly inspired by a surprisingly simple formula for MHV graviton amplitudes found by Hodges \cite{Hodges:2012ym}.

The new formulas show a deep connection between the scattering amplitudes of gluons and those of gravitons. Such a connection has been know since the Kawai-Lewellen-Tye (KLT) relations were proposed \cite{Kawai:1985xq}. These relations express graviton amplitudes as the ``square" of gauge theory amplitudes. This subject has received renewed attention in recent years due to the work of Bern, Carrasco and Johansson (BCJ) \cite{Bern:2008qj,Bern:2010ue}. The BCJ double-copy formula is a prescription to express gravity amplitudes as the ``square" of gauge theory amplitudes based on trivalent graphs. It requires that gauge theory amplitudes satisfy a type of duality between colour and kinematics that is not manifest in the Lagrangian. Therefore, it tells us something about gauge theory itself, not just gravity. The loop-level BCJ proposal remains conjectural.

The goal of this work is to explore the scattering equations which underlie the new formulas for gluons and gravitons, and to relate them to the BCJ colour-kinematics duality. We will identify and employ algebraic structures arising from the kinematics in order to make the BCJ duality manifest. Our inspiration is the case where the duality is better understood, self-dual gauge theory. As shown by the authors in \cite{Monteiro:2011pc}, the duality follows from the fact that the cubic vertex of the self-dual theory has a kinematic dependence given by the structure constant of a Lie algebra, mirroring the colour dependence. We will see that algebraic structures analogous to those of (anti-)self-dual gauge theory can be associated to each solution of the scattering equations. In particular, we will see that we have an extension of the concepts of self-dual and anti-self-dual vertices. This offers a glimpse into the loop-level generalisation of the new formulas. Recently, it has been shown that certain one-loop amplitudes can be described by these algebraic structures \cite{Boels:2013bi,Bjerrum-Bohr:2013iza}.

Since the new formulas of Cachazo, He and Yuan are very reminiscent of string theory amplitudes, it is not surprising that some objects analogous to those we will deal with have already made an appearance in that context, especially in recent work with the pure spinor formalism. Using this formalism, a complete set of tree-level BCJ numerators (which make the colour-kinematics duality manifest) was presented in \cite{Mafra:2011kj}. Signs of the BCJ duality and of the scattering equations closely related to our discussion are found in the one-loop structure of superstring amplitudes \cite{Mafra:2012kh}. The scattering equations were also identified and explored in the recent works \cite{Stieberger:2013hza,Stieberger:2013nha}, where a remarkable correspondence between superstring and supergravity amplitudes was unveiled. Another stringy context in which the scattering equations arise, rather surprisingly, is in the high energy limit of closed string amplitudes \cite{Gross:1987ar}, as pointed out in \cite{Cachazo:2013gna}.

While this work was in progress, the papers \cite{Cachazo:2013iea} and \cite{Litsey:2013jfa} appeared, showing (among other things) that it is possible to obtain BCJ numerators for each of the contributions to the gauge theory amplitude corresponding to a solution of the scattering equations. We see this result as a direct consequence of the fact, proven in \cite{Cachazo:2012uq}, that the Parke-Taylor factors arising from each solution of the scattering equations (to be reviewed next) satisfy the so-called BCJ relations. Explicit constructions to invert the linear problem and generate BCJ numerators for amplitudes satisfying the BCJ relations were presented in \cite{BjerrumBohr:2010hn,BjerrumBohr:2012mg,Boels:2012sy}. In this work, we provide a canonical set of BCJ numerators which can be constructed directly from the vertex configuration of a trivalent graph, based on algebraic structures naturally associated to the scattering equations. We hope that this construction and its connections to the notions of (anti-)self-duality will provide a new insight into the scattering equations.

This paper is organised as follows. In Section~\ref{sec:SE}, we will review the scattering equations and the new formulas for the scattering amplitudes of gravitons, gluons, and also massless double-coloured scalars, which fit into the same picture \cite{Cachazo:2013iea}. In Section~\ref{sec:review} we will review the colour-kinematics duality. A review of Ref.~\cite{Monteiro:2011pc}, which is our starting point, and the extension of the algebraic structure discussed there will be presented in Section~\ref{sec:selfdual}. The main result of this paper, the construction of manifestly BCJ-dual amplitudes, will be presented in Section~\ref{sec:numerators}. In Section~\ref{sec:gravity}, we will give expressions for gravity amplitudes in terms of trivalent graphs and in terms of the kinematic analogues of colour traces. The final discussion is in Section~\ref{sec:disc}.

%%%%%%%%%%%%%%%%%%%%%%%%%%%%%%%%%%%%%
%%%%%%%%%%%%%%%%%%%%%%%%%%%%%%%%%%%%%

\section{Review of the scattering equations}
\label{sec:SE}

The scattering equations at $n$ points are a deceptively simple set of equations, to be solved for $n$ complex variables $\sigma_a$:
\begin{equation}
\label{eq:SE}
\sum_{b \neq a} \frac{k_a \cdot k_b}{\sigma_a - \sigma_b} = 0,
\end{equation}
where the $k_a$ are the momenta of the external particles. Thus, there are $n$ equations. However, only $n - 3$ of these equations are independent.

The significance of these equations to the tree-level scattering of massless particles has recently been explored by Cachazo, He and Yuan~\cite{Cachazo:2013gna,Cachazo:2013hca,Cachazo:2013iea}. A first observation about these equations is that they are invariant under $SL(2, \mathbb{C})$ when the $k_a$ satisfy conservation of momentum. That is, given a particular solution for the $\sigma_a$, an equally good solution is given by
\begin{equation}
\sigma'_a = \frac{ A \sigma_a + B}{ C \sigma_a + D},
\end{equation}
when $A D  - B C = 1$. Thus we may interpret the $\sigma_a$ as points on $S^2$. 

Up to this redundancy, there are always $(n - 3)!$ different solutions of the scattering equations. Of course, the case $n = 3$ is trivial: one may exploit the $SL(2, \mathbb{C})$ to fix the locations of the three points; meanwhile, all the kinematic invariants vanish. For $n = 4$, after fixing $SL(2, \mathbb{C})$, there is only one variable. Meanwhile, there is one equation to be solved, which is linear in the remaining variable. Solutions of the scattering equations at $n$ points may be obtained from the solutions at $n - 1$ points according to an algorithm described in \cite{Cachazo:2013gna}; using this algorithm, each individual solution for the $(n - 1)$-point system yields $n - 3$ solutions for the $n$-point system, for a total of $(n - 3)!$ solutions.

The complete tree-level S-matrices of gauge theory and gravity in $D$ dimensions are easily written down using the solutions of the scattering equations. The colour-ordered $n$ point Yang-Mills amplitudes $A_n$, and the $n$-point gravity amplitudes $\mathcal{M}_n$ are simply \cite{Cachazo:2013gna,Cachazo:2013hca}
\begin{align}
\label{eq:intAmplitudesgauge}
A_n &= \int \frac{d^n \sigma}{\mathrm{vol \; SL}(2, \mathbb{C})} \prod_a{}' \delta \left( \sum_{b \neq a} \frac{k_a \cdot k_b}{\sigma_a - \sigma_b} \right) \frac{E_n(\{ k, \epsilon, \sigma \})}{\sigma_{12} \sigma_{23} \cdots \sigma_{n1}}, \\
\mathcal{M}_n &= \int \frac{d^n \sigma}{\mathrm{vol \; SL}(2, \mathbb{C})} \prod_a{}' \delta \left( \sum_{b \neq a} \frac{k_a \cdot k_b}{\sigma_a - \sigma_b} \right) E_n(\{ k, \epsilon, \sigma \})^2,
\label{eq:intAmplitudesgrav}
\end{align}
where $\sigma_{ab} = \sigma_a - \sigma_b$, and
\begin{equation}
\label{eq:deltafunc}
\prod_a{}' \delta \left( \sum_{b \neq a} \frac{k_a \cdot k_b}{\sigma_a - \sigma_b} \right) = \sigma_{ij} \sigma_{jk} \sigma_{ki} \prod_{a \neq i, j, k} \delta \left( \sum_{b \neq a} \frac{k_a \cdot k_b}{\sigma_a - \sigma_b} \right).
\end{equation}
This is independent of the choice of $i, j$ and $k$; therefore, it is permutation symmetric. Meanwhile, the object $E_n(\{ k, \epsilon, \sigma \})$ is a gauge-invariant function of the polarisations $\epsilon_a$ of the particles. Moreover, it is symmetric under permutations of the particles. It is most easily described in terms of an antisymmetric $2n \times 2n$ matrix $\Psi$, which is given in block form as
\begin{equation}
\Psi = \begin{pmatrix}
A & -C^T \\
C & B
\end{pmatrix}.
\end{equation}
The $n \times n$ blocks of $\Psi$ are defined by
\begin{align}
A_{ab} &= \begin{dcases}
\frac{k_a \cdot k_b}{\sigma_{ab}} & a \neq b, \\
0 & a = b,
\end{dcases} \\
B_{ab} &=  \begin{dcases}
\frac{\epsilon_a \cdot \epsilon_b}{\sigma_{ab}} & a \neq b, \\
0 & a = b,
\end{dcases}
\\
C_{ab} &=  \begin{dcases}
\frac{\epsilon_a \cdot k_b}{\sigma_{ab}} & a \neq b, \\
- \sum_{c \neq a} \frac{\epsilon_a \cdot k_c}{\sigma_{ac}} & a = b.
\end{dcases}
\end{align}
The Pfaffian of $\Psi$ vanishes; however, leaving out two rows $i$ and $j$ as well as the corresponding columns $i$ and $j$ yields a matrix $\Psi^{ij}_{ij}$ which has a non-vanishing Pfaffian. Indeed, the object $E_n(\{ k, \epsilon, \sigma \})$ appearing in the amplitudes~\eqref{eq:intAmplitudesgauge} and \eqref{eq:intAmplitudesgrav} is 
\begin{equation}
\label{eq:pfPsi}
E_n(\{ k, \epsilon, \sigma \}) \equiv \text{Pf}\,' (\Psi^{ij}_{ij}) \equiv 2\frac{(-1)^{i+j}}{\sigma_{ij}} \text{Pf} \, (\Psi^{ij}_{ij}). 
\end{equation}

The delta functions appearing under the integral sign in the amplitudes~\eqref{eq:intAmplitudesgauge} and \eqref{eq:intAmplitudesgrav} completely localise the integrals. So, in fact, there are no integrations to do, and the amplitudes can be expressed as a sum over the $(n -3 )!$ solutions of the scattering equations. A Jacobian occurs on integrating over the delta functions. To describe this Jacobian, we introduce a matrix $\Phi$ with components:
\begin{equation}
\Phi_{ab} = 
\begin{dcases}
\frac{k_a \cdot k_b}{\sigma_{ab}^2} & a \neq b, \\
 - \sum_{c \neq a} \frac{k_a \cdot k_c}{\sigma_{ac}^2} & a = b.
\end{dcases}
\end{equation}
This matrix, introduced in \cite{Cachazo:2012da}, is closely connected to the Hodges formula for MHV graviton amplitudes \cite{Hodges:2012ym}.
Because the delta functions \eqref{eq:deltafunc} instruct us to omit rows $i, j$ and $k$, we omit these from the Jacobian determinant. In addition, to gauge fix the $SL(2, \mathbb{C})$ we may fix the position of three points, say $\sigma_r, \sigma_s$ and $\sigma_t$. As usual, this gauge-fixing procedure introduces a factor $\sigma_{rs} \sigma_{st} \sigma_{ts}$ into the integral. The result is that the Jacobian determinant is the minor determinant of $\Phi$, omitting rows $i, j$ and $k$ and columns $r, s$ and $t$. It is convenient to introduce the simple notation
\begin{equation}
\label{eq:detPhi}
\text{det}\, ' \Phi = \frac{|\Phi|^{ijk}_{rst}}{\sigma_{rs} \sigma_{st} \sigma_{ts} \sigma_{ij} \sigma_{jk} \sigma_{ki}}.
\end{equation}
The amplitudes are then written as sums over contributions from distinct solutions of the scattering equations,
\begin{align}
\label{eq:amplitudegauge}
A_n &= \sum_{\text{solutions}} \frac{1}{\sigma_{12} \sigma_{23} \cdots \sigma_{n1}} \frac{\text{Pf} \, ' \Psi}{\text{det}\,' \Phi} , \\
% \frac{\sigma_{rs} \sigma_{st} \sigma_{ts} \sigma_{ij} \sigma_{jk} \sigma_{ki}}{|\Phi|^{ijk}_{rst}}, \\
\mathcal{M}_n &= \sum_{\text{solutions}}  \frac{(\text{Pf} \, ' \Psi)^2}{\text{det}\,' \Phi}.
% E_n(\{ k, \epsilon, \sigma \})^2 \frac{\sigma_{rs} \sigma_{st} \sigma_{ts} \sigma_{ij} \sigma_{jk} \sigma_{ki}}{|\Phi|^{ijk}_{rst}},
\label{eq:amplitudegrav}
\end{align}

The scattering equations are relevant not just for the scattering of spin 1 and spin 2 particles, but also for the scattering of certain massless scalar theories. These are theories of scalar fields $\phi^{aa'}$ with a double-coloured cubic vertex, where $a$ and $a'$ transform under the adjoint of two groups $G$ and $G'$. Ref.~\cite{Cachazo:2013iea} presents an expression for the amplitudes of particles of spin $\bf s$, valid for these scalar theories as well as for gluons and gravitons,
\begin{equation}
\label{eq:amplitudespins}
\mathcal{A}_n^{({\bf s})} = \sum_{\text{solutions}}
\left( \frac{\text{Tr}(T^{a_1}T^{a_2}\cdots T^{a_n})}{\sigma_{12} \sigma_{23} \cdots \sigma_{n1}} 
+\text{non-cyclic permutations}\right)^{2-\bf s}
\frac{(\text{Pf} \, ' \Psi)^{\bf s}}{\text{det}\,' \Phi}.
\end{equation}
In the scalar case, we can have distinct groups $G\neq G'$, or simply distinct algebra indices $a_r\neq a'_r$, so that we should substitute
\begin{equation}
\left(\frac{\text{Tr}(T^{a_1}T^{a_2}\cdots T^{a_n})}{\sigma_{12} \sigma_{23} \cdots \sigma_{n1}} + \ldots\right)^{2} \to
\left(\frac{\text{Tr}(T^{a_1}T^{a_2}\cdots T^{a_n})}{\sigma_{12} \sigma_{23} \cdots \sigma_{n1}} + \ldots\right)
\left(\frac{\text{Tr}(T^{a'_1}T^{a'_2}\cdots T^{a'_n})}{\sigma_{12} \sigma_{23} \cdots \sigma_{n1}} + \ldots\right).
\end{equation}

The most remarkable property of the solutions of the scattering equations, which leads to the simplicity of the formulas above, is the so-called KLT orthogonality, discovered in \cite{Cachazo:2012da} and proven in \cite{Cachazo:2013gna}. The KLT relations \cite{Kawai:1985xq,Bern:1998sv}, proven in \cite{BjerrumBohr:2010ta}, give a graviton amplitude as a product of two sets of gauge theory colour-ordered amplitudes, 
\begin{equation}
\label{eq:KLT}
\mathcal{M}_n = \sum_{P,P'\in S_n}
A_n(P) S_n^{\text{KLT}}(P,P') A_n(P'),
\end{equation}
mediated by a momentum kernel $S_n^{\text{KLT}}$ dependent on the Mandelstam variables \cite{BjerrumBohr:2010hn}. This defines a natural inner product (equivalent to the BCJ double-copy to be reviewed later). The statement of KLT orthogonality is that the Parke-Taylor amplitudes constructed from solutions of the scattering equations, e.g.
\begin{equation}
\frac{1}{\sigma_{12} \sigma_{23} \cdots \sigma_{n1}},
\end{equation}
are orthogonal with respect to the KLT inner product when they arise from two different solutions. To be more specific, in the KLT product above, let $A_n(P)$ denote the permutations of a Parke-Taylor factor with solution $I$ of the scattering equations, and let $A_n(P')$ denote the permutations of a Parke-Taylor factor with solution $J\neq I$; then, the product vanishes. The consequence of this fact is that the expression for gravity amplitudes \eqref{eq:amplitudegrav} contains only one sum over solutions of the scattering equations, and not two sums with mixed contributions, as would in principle arise from the KLT relations.

%%%%%%%%%%%%%%%%%%%%%%%%%%%%%%%%%%%%%
%%%%%%%%%%%%%%%%%%%%%%%%%%%%%%%%%%%%%

\section{Review of the colour-kinematics duality}
\label{sec:review}

There are two main aspects to the BCJ story: colour-kinematics duality, and the double-copy relation between gauge and gravity amplitudes. The duality and the double-copy were first noticed by Bern, Carrasco and Johansson at tree level~\cite{Bern:2008qj}, and were later generalised by the same authors to loops~\cite{Bern:2010ue}. Our focus in this paper will be on tree level, and therefore we will restrict our review to this case for simplicity.

We begin by discussing colour-kinematics duality. This is a property of amplitudes in gauge theory (with or without supersymmetry). It is always possible to express gauge amplitudes as a sum over cubic diagrams. One way to achieve this is to begin with Feynman diagrams, and then to systematically assign diagrams with four point vertices to cubic diagrams by introducing the missing propagator denominators with a compensating factor in the numerator. For example, 
\begin{equation}
\begin{minipage}[c]{0.2\linewidth}
\centering
\includegraphics[scale=0.3]{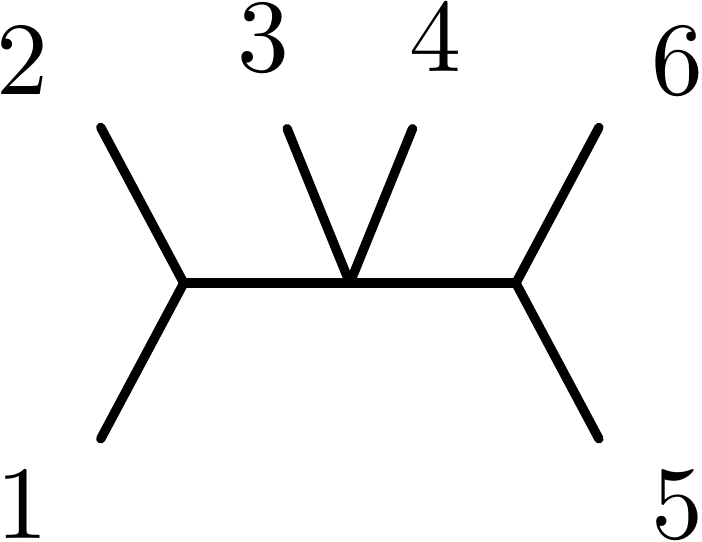}
\end{minipage}
=
\begin{minipage}[c]{0.15\linewidth}
\centering
\includegraphics[scale=0.3]{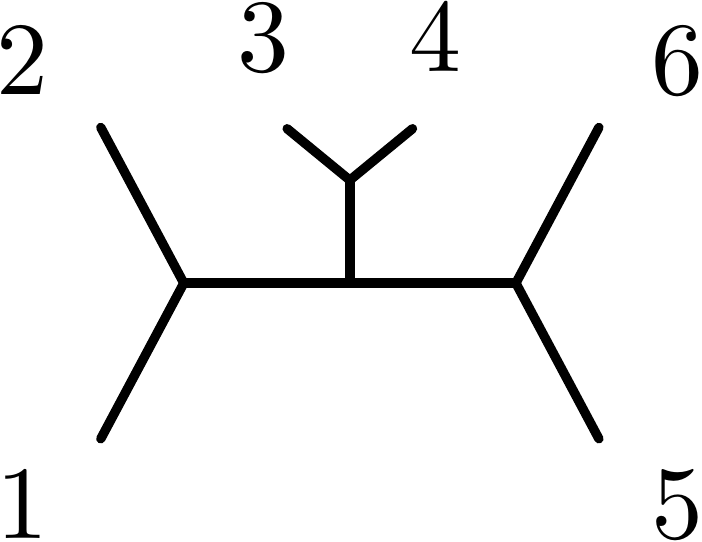}
\end{minipage}
s_{34} \;\;
=
\begin{minipage}[c]{0.15\linewidth}
\centering
\includegraphics[scale=0.3]{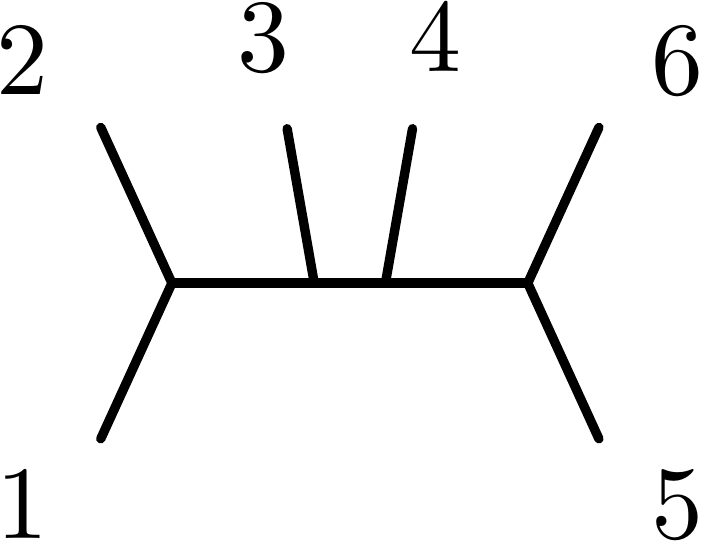}
\end{minipage}
s_{123}.
\end{equation}
We see that this is not unique. In gauge theory, diagrams are associated not only with a set of propagators and a kinematic numerator, but also with a colour factor. Since the colour factors are built from the structure constants $\tf^{abc}$ of the gauge group\footnote{Following \cite{Bern:2008qj,Bern:2010ue}, we use a tilde to specify the normalisation $\tf^{abc}=\textrm{Tr}([T^a,T^b]T^c)$.}, they are cubic in nature. Thus, we assign contact terms to cubic diagrams by inspecting their colour structure, and introducing the missing propagators with compensating factors in the numerators.

We may therefore express any (colour-dressed) $n$-point amplitude in a gauge theory as
\begin{equation}
\mathcal{A}_n = \sum_{\alpha \in \mathrm{cubic}} \frac{n_\alpha c_\alpha}{D_\alpha},
\end{equation}
where the sum runs over the set of distinct cubic $n$-point diagrams. Meanwhile, the objects $n_\alpha$, $c_\alpha$ and $D_\alpha$ are the kinematic numerators, colour factors and (total) propagator denominators associated with the diagram $\alpha$. The statement of colour-kinematics duality is now simple. Take any triple $(\alpha, \beta, \gamma)$ of diagrams such that their colour factors are related by a Jacobi identity:
\begin{equation}
c_\alpha + c_\beta + c_\gamma = 0.
\end{equation}
It is possible to find a set of valid numerators such that
\begin{equation}
n_\alpha + n_\beta + n_\gamma = 0.
\end{equation}
Moreover, these kinematic numerators have the same antisymmetry properties as the colour factors. If, under interchanging two legs, a colour factor changes sign, then so does the corresponding kinematic factor:
\begin{equation}
c_\alpha \rightarrow - c_\alpha \Rightarrow n_\alpha \rightarrow - n_\alpha.
\end{equation}
Thus, the kinematic structure mirrors the algebraic structure of the colour factors. In general, such choice of numerators is non-unique, but non-trivial to find. By now, various algorithms have been described for finding BCJ numerators; see, for example, \cite{BjerrumBohr:2010hn,Mafra:2011kj,BjerrumBohr:2012mg,Boels:2012sy,Fu:2012uy}. We shall discuss below how the scattering equations are naturally associated with a certain set of numerators.

Before we move on, let us dwell a little on one previous algorithm for finding BCJ numerators. In fact, the algorithm proposed in~\cite{BjerrumBohr:2012mg} is particularly relevant for understanding the scattering equations, as we shall see. The idea of~\cite{BjerrumBohr:2012mg} is simple. There are $(n-3)!$ linearly independent colour-ordered gauge amplitudes for $n$ points~\cite{Bern:2008qj,BjerrumBohr:2009rd,Stieberger:2009hq,Feng:2010my}. Thus, we can view the colour-ordered amplitudes as a vector $\vec A$ in an $(n-3)!$-dimensional space. It is straightforward to construct another vector in the same space. Indeed, consider a theory involving a set of massless scalar fields $\phi^{aa'}$ where $a$ and $a'$ transform under the adjoint of two groups $G$ and $G'$. We suppose that these scalar fields interact only through a cubic vertex, with Feynman rule $\tf^{abc} \tf^{a'b'c'}$. These are exactly the scalar theories considered in the previous section. We may then construct colour-ordered amplitudes $\vec \theta$ for this theory, with respect to the group $G$, say. Because the analogues of the kinematic numerators in this scalar theory are, in fact, directly made of the structure constants of the group $G'$, they automatically satisfy colour-kinematics duality, so that there are $(n-3)!$ independent amplitudes. In short, $\vec \theta$ is an element of the same vector space as the gauge amplitudes $\vec A$. Since there are infinitely many groups to choose from (and different particle labellings within each group), we can find many vectors living in this space. Indeed we may find a basis of this space using the scalar amplitudes. Therefore, the gauge amplitudes $\vec A$ are a linear combination of the scalar amplitudes:
\begin{equation}
\label{eq:linearBCJ}
\vec A = \sum_{I=1}^{(n-3)!} \alpha^{(I)} \vec \theta^{(I)},
\end{equation}
where $\vec \theta^{(I)}$ is the vector of independent colour-ordered amplitudes of the $I$th scalar theory.
Now, consider the $k$th colour-ordered gauge amplitude; denoting the numerator of the scalar theory by $c'$, it is given by
\begin{equation}
A_k = \sum_{I=1}^{(n-3)!} \alpha^{(I)} \sum_{\alpha \in k\mathrm{th \; ordered \; cubic}} \frac{c'^{(I)}_\alpha}{D_\alpha}
=\sum_{\alpha \in k\mathrm{th \; ordered \; cubic}} \frac{1}{D_\alpha} 
\left(\sum_{I=1}^{(n-3)!} \alpha^{(I)} c'^{(I)}_\alpha\right).
\end{equation}
The $\alpha$-sum runs over the cubic diagrams in the $k$th colour order. Notice that the gauge numerators have been expressed as a linear combination of the $c'^{(I)}$; these automatically satisfy Jacobi identities because they are built from the structure constants $\tf^{a'b'c'}$ of the group $G'$. In this way, we can always find colour-dual kinematic numerators. However, to do so we must invert the linear system \eqref{eq:linearBCJ} to compute the expansion coefficients $\alpha^{(I)}$. We will see that the work of Cachazo, He and Yuan \cite{Cachazo:2013gna,Cachazo:2013hca} provides a canonical choice for the basis amplitudes, and an explicit formula for the $\alpha^{(I)}$. In particular, the basis is orthogonal with respect to the natural inner product, the double-copy formula, which we briefly review below.

One intriguing aspect of colour-kinematics duality is that it suggests that there exists an algebraic structure which can be used to directly compute BCJ numerators in gauge theories, including at loop level, using a procedure analogous to Feynman rules. However, to date an understanding of this algebraic structure remains elusive, except in the special case of self-dual Yang-Mills theory in four dimensions~\cite{Monteiro:2011pc}. In the self-dual theory, it is known that the kinematic algebra is an area-preserving diffeomorphism algebra. It was also shown in~\cite{Monteiro:2011pc} that colour-dual kinematic numerators for MHV amplitudes in (full) Yang-Mills theory can be computed directly from a knowledge of the self-dual theory, using a particular choice of gauge. As we shall see below, the scattering equations allow us to extend essentially the same structure to the full theory, in any dimension and for any polarisations of the external particles. Since this material is central to this paper, we will review it in detail in Section~\ref{sec:selfdual}.

The second major aspect of the BCJ story~\cite{Bern:2008qj,Bern:2010ue} is the double-copy formula which relates gauge amplitudes and gravitational amplitudes. Given a set of colour-dual kinematic numerators $n_\alpha$, valid for an $n$-point gauge amplitude, there is an associated gravitational amplitude\footnote{We are fixing the normalisation of gravitational amplitudes according to \eqref{eq:amplitudegrav}.}
\begin{equation}
(-2)^{n-3} \mathcal{M}_n = \sum_\alpha \frac{n_\alpha n_\alpha}{D_\alpha}.
\end{equation}
That is, the gravitational amplitude is obtained from the gauge amplitude by replacing the colour factors of the gauge amplitude by another copy of the kinematic numerators. More generally, we may build a gravitational amplitude using $n$-point numerators of different gauge theories. For example, we may construct a gravitational amplitude from one set of pure Yang-Mills numerators $n_\alpha$ and a set of $\mathcal{N} = 4$ super-Yang-Mills numerators $\tilde n_\alpha$. The result is an amplitude in $\mathcal{N} = 4$ supergravity, given by
\begin{equation}
(-2)^{n-3} \mathcal{M}_n = \sum_{\alpha \in \mathrm{cubic}} \frac{n_\alpha \tilde n_\alpha}{D_\alpha}.
\end{equation}
%The fact that the kinematic numerators appear in gravity indicates that the physics of BCJ is gravitational---this is reflected in the fact that the relevant algebra for the MHV amplitudes is a diffeomorphism algebra.
In fact, only one of the numerators, $n_\alpha$ or $\tilde n_\alpha$, needs to satisfy the colour-kinematics duality.
The double-copy formula has been proven at tree level~\cite{Bern:2010yg,Cachazo:2013iea}. This procedure is therefore equivalent to the KLT relations.

At loop level, the colour-kinematics duality remains conjectural. However, it has been verified in various examples \cite{Carrasco:2011mn,Boels:2011tp,Bern:2012uf,Carrasco:2012ca,Boels:2012ew,Oxburgh:2012zr,Bjerrum-Bohr:2013iza,Bern:2013yya,Boels:2013bi,Nohle:2013bfa}.

%%%%%%%%%%%%%%%%%%%%%%%%%%%%%%%%%%%%%%%%
%%%%%%%%%%%%%%%%%%%%%%%%%%%%%%%%%%%%%%%%

\section{Self-duality and the scattering equations}
\label{sec:selfdual}

We will start by reviewing the kinematic algebras arising in (anti)-self-dual gauge theory, and their relation to the colour-kinematics duality \cite{Monteiro:2011pc}. Then we will see that an extension of these algebras applies more generally in the context of the scattering equations.

\subsection{Self-dual and anti-self-dual sectors in four dimensions}

Consider four-dimensional gauge theory in light-cone gauge, with light-cone coordinates $(u,v,w,\wb)$. We will follow the discussion in \cite{Monteiro:2011pc}. The spacetime metric is defined such that, for any two four-vectors $A$ and $B$,
\be
2A \cdot B=A_u B_v + A_v B_u - A_w B_{\wb} - A_{\wb} B_w.
\ee
We take the polarisation vectors to be
\begin{align}
\label{polpm}
\varepsilon_a^{+} = (0,{k_a}_w,0,{k_a}_u), \qquad \varepsilon_a^{-} = (0,{k_a}_{\wb},{k_a}_u,0),
\end{align}
so that they satisfy
\be
\label{polpmortho}
\varepsilon_a^{\pm} \cdot \varepsilon_b^{\pm} = 0, \qquad 2 \varepsilon_a^{+} \cdot \varepsilon_b^{-} = - {k_a}_u {k_b}_u.
\ee
The contraction of the polarisation vectors with the momenta determines two antisymmetric bilinear forms of interest,
\begin{align}
2\varepsilon_a^{+}\cdot k_b &= {k_a}_w {k_b}_u -  {k_a}_u {k_b}_w \equiv X_{a,b}, 
\label{eps4dp}\\
2\varepsilon_a^{-}\cdot k_b &= {k_a}_{\wb} {k_b}_u - {k_a}_u {k_b}_{\wb} \equiv \Xb_{a,b},
\label{eps4dm}
\end{align}
and we have
\be
\label{sabXXbar}
s_{ab} = \frac{X_{a,b}\Xb_{a,b}}{{k_a}_u {k_b}_u}.
\ee
While the latter equation is valid only if $k_a$ and $k_b$ are on-shell, we can extend the definitions \eqref{eps4dp}-\eqref{eps4dm} to allow for momenta $k$ which are not massless; for instance $X_{1+2,3}=X_{1,3}+X_{2,3}$. For clarity, we will use capital Greek letters $A,B,\ldots$ to denote off-shell momenta later on.

Let us connect these quantities to the scattering equations. We write a four-vector with spinorial indices as
\be
k_{\alpha \dalpha} = \left(  \begin{array}{cc} {k}_u & {k}_w \\ {k}_{\wb} & {k}_v 
\end{array} \right).
\ee
If the four-vector is on-shell, then this matrix is singular, and it can be written in terms of two spinors as $k_{\alpha \dalpha} = \lambda_\alpha {\tlambda}_{\dalpha} $. Let us choose these spinors to be
\be
\lambda_\alpha = \left(  \begin{array}{c} 1 \\ \sigma \end{array} \right), \qquad 
{\tlambda}_{\dalpha} = \left(  \begin{array}{c} k_u \\ k_w \end{array} \right),
\qquad \textrm{where} \quad \sigma = \frac{k_{\wb}}{k_u}=\frac{k_v}{k_w}.
\ee
We can now define the spinor products ($\epsilon^{12}=-1$)
\begin{align}
\label{sec1Xbar}
\langle ab \rangle &= \epsilon^{\alpha \beta} \lambda^{(a)}_\alpha \lambda^{(b)}_\beta = \sigma_a - \sigma_b = \frac{\Xb_{a,b}}{{k_a}_u {k_b}_u}, \\
[ ab ] &= - \epsilon^{\dalpha \dbeta}  \tlambda^{(a)}_{\dalpha} \tlambda^{(b)}_{\dbeta} =- X_{a,b}.
\end{align}
From the relation \eqref{sabXXbar}, or equivalently $s_{ab}=\langle ab \rangle[ba]$, we obtain
\be
\label{sec1X}
X_{a,b} = \frac{s_{ab}}{\sigma_a - \sigma_b}.
\ee
Notice that, from the definitions \eqref{eps4dp}-\eqref{eps4dm} and from momentum conservation, we have
\be
\sum_{b\neq a} X_{a,b} = 0, \qquad  \sum_{b\neq a} \Xb_{a,b} = 0.
\ee
These identities take the form of the scattering equations \eqref{eq:SE}. The identity on the left corresponds to $\sigma_a = {k_a}_{\wb}/{k_a}_u$, as considered above, while the identity on the right can be seen as the equivalent statement for its conjugate $\bar{\sigma}_a = {k_a}_w/{k_a}_u$. So, in four dimensions and for $n>4$, these are always two of the solutions to the scattering equations. In particular, these two solutions give the single non-vanishing contribution to the MHV and the $\overline{\textrm{MHV}}$ amplitudes, respectively; the factor $\text{Pf}\;'\Psi$ in the expression \eqref{eq:amplitudegauge} vanishes for the other solutions with such a choice of polarisations. The $n=4$ case, where MHV$=\overline{\textrm{MHV}}$, is special: there is a single solution to the scattering equations, and indeed the two solutions discussed above are the same, up to an $SL(2,\Co)$ transformation. To see this, let us fix the $SL(2,\Co)$ freedom by performing a transformation such that $(\sigma_1,\sigma_2,\sigma_3)\to (\infty,0,1)$. Then
\be
\sigma_4 \to {\sigma'}_4 = \frac{(\sigma_1-\sigma_3)(\sigma_2-\sigma_4)}{(\sigma_1-\sigma_4)(\sigma_2-\sigma_3)}.
\ee
The cross ratio on the right-hand-side is equal to its conjugate ($\bar{\sigma}$ replacing $\sigma$), since
\be
\frac{(\sigma_1-\sigma_3)(\sigma_2-\sigma_4)}{(\sigma_1-\sigma_4)(\sigma_2-\sigma_3)} =
\frac{s_{13}}{s_{14}} \frac{X_{1,4}(\sigma_2-\sigma_4)}{X_{1,3}(\sigma_2-\sigma_3)} =
 - \frac{s_{13}}{s_{14}},
\ee
where we used momentum conservation for the second equality, $X_{1,3}(\sigma_2-\sigma_3) =- X_{1,4}(\sigma_2-\sigma_4)$.

The action of $SL(2,\Co)$ on this type of solution to the scattering equations,
\be
\sigma \to  {\sigma'} = \frac{A \sigma + B}{C \sigma + D}, \quad AD - BC =1,
\ee
corresponds to the standard action of $SL(2,\Co)$ on the spinor $\lambda$,
\be
\lambda_\alpha \to  {\lambda'}_\alpha = 
\left(  \begin{array}{cc} D & C \\ B & A \end{array} \right)  \left(  \begin{array}{c} 1 \\ \sigma \end{array} \right)
= t\left(  \begin{array}{c} 1 \\ \sigma' \end{array} \right),
\qquad t= C \sigma + D,
\ee
followed by a rescaling of the other spinor, $\tlambda_{\dalpha} \to {\tlambda'}_{\dalpha} = t^{-1} \tlambda_{\dalpha}$.

Our interest in the quantity $X_{A,B}$ arises from the fact that it is the kinematic part of the vertex of self-dual gauge theory \cite{Cangemi:1996rx,Chalmers:1996rq}:
\vspace{-2.5cm}
\begin{equation}
\label{3ptvertex}
\setlength{\unitlength}{2cm}
\begin{picture}(2,2)
\put(-1.5,0){\line(-1,1){0.6}}
\put(-1.5,0){\line(-1,-1){0.6}}
\put(-1.5,0){\line(1,0){0.849}}
\put(0.2,0){$=\; (2 \pi)^4 \delta^{(4)}(K_A+K_B +K_C)\, X_{A,B}\, f^{a_A a_B a_C} .$}
\put(-2.3,0.6){$A$}
\put(-2.3,-0.7){$B$}
\put(-0.5,0.0){$C$}
\end{picture}
\end{equation}
\vspace{0.8cm}

\noindent Seeing the vertex as part of a trivalent graph, $K_A,K_B,K_C$ correspond to sums of momenta of the external particles; for instance, if $K_A=k_1+k_2$ and $K_B=k_3$, then $X_{A,B}=X_{1+2,3}= X_{1,3}+X_{2,3}$. A basic condition for $X_{A,B}$ to represent a vertex is that we can read it with any two of the three legs meeting at that vertex, e.g. $X_{A,B}=X_{B,C}=X_{C,A}$ in the figure above. This is ensured by momentum conservation, and by the fact that $X$ is an antisymmetric bilinear form.

Next, we review how this vertex is related to the colour-kinematics duality, as shown in \cite{Monteiro:2011pc}. The main point is that the vertex $X_{A,B}$ is the structure constant of a Lie algebra. In particular, it is the Lie algebra of area-preserving diffeomorphisms in the plane $w-u$, generated by the vectors
\be
\label{eq:vpol}
V^+_A = -2e^{-iK_A\cdot x} \epsilon_A^+ \cdot \partial \qquad \textrm{such that} \qquad [V^+_A,V^+_B]= iX_{A,B}V^+_{A+B}.
\ee
We take the definition \eqref{polpm} of the polarisation vectors to extend to off-shell momenta. The Jacobi identity involving the generators leads to
\be
X_{A,B}X_{A+B,C}+X_{B,C}X_{B+C,A}+X_{C,A}X_{C+A,B}=0.
\ee
Therefore, in self-dual gauge theory, we have a single cubic vertex whose kinematic part satisfies Jacobi identities, exactly like the colour part. This is the simplest manifestation of the colour-kinematics duality, since the duality is satisfied already by Feynman diagrams.

Now we can write BCJ numerators for self-dual gauge theory. Consider, for instance, the four-point case. The numerators corresponding to the three channels, associated to the colour factors
\be
\tf^{a_1 a_2 b}\tf^{b a_3 a_4}, \quad \tf^{a_2 a_3 b}\tf^{b a_1 a_4} , \quad \tf^{a_3 a_1 b}\tf^{b a_2 a_4} ,
\ee
are given by
\be
n_{12,34} = \alpha X_{1,2}X_{3,4}, \quad n_{23,14} = \alpha X_{2,3}X_{1,4}, \quad n_{31,24} = \alpha X_{3,1}X_{2,4},
\ee
respectively. We introduced a factor $\alpha$ which is independent of the particle ordering, and which takes into account overall factors coming from polarisations and their normalisation; we are not interested in this factor for now. The construction of numerators in general is straightforward. To give a more elaborate example, consider the seven-point trivalent graph
\be
\tf^{a_1 a_2 b}\tf^{b a_3 c}\tf^{c d e}\tf^{d a_4 a_5} \tf^{e a_6 a_7} ,
\ee
whose numerator is 
\be
n = \alpha X_{1,2}X_{1+2,3}X_{4+5,6+7}X_{4,5}X_{6,7}.
\ee

We described how to find BCJ numerators for self-dual gauge theory. However, the tree-level scattering amplitudes of self-dual gauge theory vanish. They correspond to helicity configurations where there is a single external particle with negative helicity, and it is well known that such amplitudes vanish by Ward identities (for $n>3$). To see that there is a single particle with negative helicity, notice that a self-dual field is generated perturbatively by positive helicity sources. In a Feynman diagram expansion of the field, the external legs are the sources and the field itself. If we take the sources to have incoming momentum, then the field has positive helicity if it is seen as outgoing. However, if we take it to be incoming, like the sources, then it has negative helicity. Later on, we will describe how to construct BCJ numerators for any tree-level amplitude from the scattering equations, by generalisation of the self-dual case.

Let us point out that the S-matrix of self-dual gauge theory is not trivial, that is, there exist non-vanishing gauge theory amplitudes which rely only on the self-dual vertex. These are the one-loop amplitudes where all the particles have positive helicity. Therefore, this infinite class of amplitudes can be easily shown to possess BCJ numerators, as discussed in \cite{Boels:2013bi}.

\subsection{Generalisation from the scattering equations}

In this section, we will extend the concept of the self-dual vertex discussed above so that it applies to a generic solution of the scattering equations, independently of the number of spacetime dimensions.

A natural observation is that solutions still come in pairs in a certain sense. If our choice of external momenta is such that the Mandelstam variables $s_{ab}$ are real-valued, then the complex conjugate of a solution to the scattering equations is clearly also a solution. However, it is not clear that one can obtain a relation analogous to \eqref{sabXXbar} from those two conjugate solutions; indeed, generically it is not possible. To be more precise, let us define these quantities for a given solution $\sigma_a$ of the scattering equations as
\be
\label{defXXbar}
X_{a,b} \equiv \frac{s_{ab}}{\sigma_a-\sigma_b}, \qquad \Xb_{a,b} \equiv (\sigma_a-\sigma_b) h_a h_b ,
\ee
and we also set $X_{a,a}\equiv 0$. We introduced the quantities $h_a$, which are the analogues of the ${p_a}_u$ from the previous section, so that the definition of $\Xb_{a,b}$ generalises the relation \eqref{sec1Xbar}. The $h_a$ must obey the constraints
\be
\label{eq:condh}
\sum_{a=1}^n  h_a = \sum_{a=1}^n  \sigma_a h_a = 0,
\ee
so that we still have
\be
\label{SEXXbar}
\sum_{b\neq a} X_{a,b} = 0, \qquad  \sum_{b\neq a} \Xb_{a,b} = 0.
\ee
We shall impose no other constraints on the $h_a$. The question of whether $X$ and $\bar{X}$ are obtained from conjugate solutions is very simple: can $\Xb_{a,b}$, as defined in $\eqref{defXXbar}$, be given by $\Xb_{a,b}=s_{ab}/(\bar{\sigma}_a-\bar{\sigma}_b)$ for a solution to the scattering equations denoted by $\bar{\sigma}_a$? To address this question, one can try to solve the equations
\be
X_{a,b} \Xb_{a,b} = s_{ab} h_a h_b
\ee
using $\Xb_{a,b}=s_{ab}/(\bar{\sigma}_a-\bar{\sigma}_b)$. We find numerically that there is no solution for $h_a$ in general, unless the pair of solutions $\sigma_a$ and $\bar{\sigma}_a$ admits a four-dimensional interpretation in terms of projective holomorphic and anti-holomorphic spinors, as in the previous section.\footnote{Notice that this is always the case at five points, since the scattering occurs in a four-dimensional subspace (due to momentum conservation). Beyond five points, it is only possible generically in four spacetime dimensions, and there is a single pair of such solutions.}
Henceforth, we will consider $\Xb_{a,b}$ without assuming that it is in any way related to solutions of the scattering equations other than the one used in its definition \eqref{defXXbar}.

The generalisation \eqref{defXXbar} of the vertices $X$ and $\bar{X}$ breaks the symmetry between them encountered in the previous section, but maintains a crucial property: the equations \eqref{SEXXbar}. We shall now see that these equations give the consistency condition for $X$ (or $\bar{X}$) to behave like a three-point vertex \eqref{3ptvertex}. Let us define the ``off-shell" version of $X$ as
\be
X_{A,B} = \sum_{a\in \{A\}} X_{a,B} = \sum_{b\in \{B\}} X_{A,b} = \sum_{a\in \{A\}} \sum_{b\in \{B\}} X_{a,b},
\ee
where $\{A\}$ and $\{B\}$ are two sets of external particles. The picture is that, for a vertex like \eqref{3ptvertex} in a tree-level graph, we have a partition of the external particles into the three sets $\{A\}$, $\{B\}$ and $\{C\}$. These sets contain the particles connected through the graph to the lines $A$, $B$ and $C$ of the vertex, respectively. Then we must have, as a consistency condition, that the vertex can be read with any two of the three lines,
\be
\label{consistent}
X_{A,B} = X_{B,C} = X_{C,A}.
\ee
This is precisely what the scattering equations guarantee. Let us see this in detail,
\begin{align}
X_{A,B} &= \sum_{a\in \{A\}} \sum_{b\in \{B\}} X_{a,b} = - \sum_{a\in \{A\}} \sum_{c\notin \{B\}} X_{a,c} = \nonumber \\
&= - \sum_{a\in \{A\}} \sum_{c\in \{A\}} X_{a,c} - \sum_{a\in \{A\}} \sum_{c\in \{C\}} X_{a,c} = - X_{A,C} = X_{C,A}.
\end{align}
The scattering equations where used in the second equality, first line. From the first to the second line, we used that the complement of $\{B\}$ is $\{A\} \cup \{C\}$. In the second line, the first term vanishes because $X_{A,A}=0$.

Consider now a partition of the external particles into four sets, $\{A\},\{B\},\{C\},\{D\}$, corresponding to the four lines involved in a Jacobi identity. The Jacobi identity for the vertices $X$ (or $\Xb$) follows directly from \eqref{consistent}, and therefore from the scattering equations:
\begin{align}
& X_{A,B}X_{C,D} + X_{B,C}X_{A,D} +X_{C,A}X_{B,D} = \nonumber \\
&= - X_{A,B}(X_{A,D}+X_{B,D}) - X_{B,C}(X_{B,D}+X_{C,D}) -X_{C,A}(X_{C,D}+X_{A,D}) =0.
\end{align}

We have defined objects $X$ and $\Xb$ which can be interpreted as the kinematic part of cubic vertices, and which obey Jacobi identities. These Jacobi identities imply that we are dealing with Lie algebras,
\be
\label{eq:hvLie}
[\hv^+_A,\hv^+_B]= iX_{A,B}\hv^+_{A+B}, \qquad [\hv^-_A,\hv^-_B]= i\Xb_{A,B}\hv^-_{A+B}.
\ee
The na\"ive extension of the representation \eqref{eq:vpol} doesn't work; there is in general no set of vectors $\hat{\varepsilon}^+_a$ such that $2\hat{\varepsilon}^+_a\cdot k_b=X_{a,b}$, mirroring \eqref{eps4dp}. Nevertheless, there are Lie algebras whose generators $\hv^\pm_A$ can be associated to lines in a trivalent graph. We will see in later sections how the elements $\hv^\pm_A$ can be used to write BCJ numerators for gauge theory amplitudes.

\subsection{Proof of the vanishing of the $X$-amplitudes}
\label{subsec:proofXvanish}

It is now natural to address the following question. The vertices $X$ from the previous section, corresponding to anti-holomorphic spinor brackets in four dimensions (and their off-shell extension) led to BCJ numerators for tree-level self-dual amplitudes. These amplitudes vanish, although the individual numerators don't. So we would like to know what happens if we consider a certain solution to the scattering equations, $\sigma_a$, and compute the ``amplitudes" constructed uniquely from the vertices $X$ associated to $\sigma_a$, which we can call the $X$-amplitudes. We find that they vanish for $n\geq4$, just like in the self-dual case. On the other hand, we find that the analogous $\Xb$-amplitudes don't vanish. (For the special four-dimensional solutions to the scattering equations, the vanishing of the $\Xb$-amplitudes requires specific values for the $h_a$, corresponding to ${p_a}_u$.)

The proof that the X-amplitudes vanish closely follows an argument given by Cangemi~\cite{Cangemi:1996rx}. First, notice that the scattering equations for $n$ particles imply that, for $m\leq n$,
\begin{equation}
\label{eq:cangemi}
\sum_{j = 1}^{m -1}  \sum_{s = 1}^j  \sum_{t = j + 1}^m X_{s,t} \sigma_{j, j+1} = (p_1 + \cdots + p_m)^2.
\end{equation}
We may prove this result by a simple rearrangement of the summation:
\begin{align}
\sum_{j = 1}^{m -1}  \sum_{s = 1}^j  \sum_{t = j + 1}^m X_{s,t} \sigma_{j, j+1} &= \sum_{1 \leq s \leq j < t \leq m} X_{s,t} \sigma_{j, j+1} \\
&=\sum_{1 \leq s < t \leq m} X_{s,t} \sigma_{s,t} \\
&= \sum_{1 \leq s < t \leq m} s_{s,t} \\
&= \left( \sum_{s = 1}^m p_s \right)^2.
\end{align}
We will begin with a simple example at four points. The colour-ordered $X$-amplitude is
\begin{equation}
A^{(X)}(1,2,3,4)=A^{(X)}_4= \frac{X_{1,2} X_{3,4}}{s_{12}} +  \frac{X_{4,1} X_{2,3}}{s_{23}} = \frac{\sigma_{23} \left( X_{1,3} + X_{2,3} \right) + \sigma_{12} \left(X_{1,2} + X_{1,3} \right)}{\sigma_{12} \sigma_{23}}.
\end{equation}
By Eq.~\eqref{eq:cangemi}, we see that we get the simple result
\begin{equation}
A^{(X)}_4 = \frac{p_4^2}{\sigma_{12} \sigma_{23}}.
\end{equation}
Thus, when $p_4$ is on shell, the amplitude vanishes. We can also consider the case $p_4^2 \neq 0$; for example, this could be a four-point subamplitude of a larger amplitude. Indeed, let us consider the five-point $X$-amplitude as a further illustration. The amplitude is
\begin{align}
A^{(X)}_5 =  &\frac{X_{1,2} X_{3,4} X_{1+2,3+4}}{s_{12} s_{34}} 
+ 
\left(  \frac{X_{1,2}X_{1+2,3}}{s_{12}} + \frac{X_{2,3} X_{1,2+3}}{s_{23}} \right) \frac{X_{4,5}}{s_{123}} \nonumber \\
&\quad+ \left(  \frac{X_{2,3} X_{2+3,4}}{s_{23}} + \frac{X_{3,4}X_{2,3+4}}{s_{34}} \right) \frac{X_{5,1}}{s_{234}} .
\end{align}
The last two terms in this expression have the structure of off-shell colour-ordered four-point subdiagrams, connected to a three-point tree involving particle $5$. Meanwhile, the first term includes two three-point vertices, which we view as off-shell colour-ordered three-point subdiagrams, connected to a three-point tree involving particle $5$. Thus,
\begin{align}
A^{(X)}_5 &= \frac{1} {\sigma_{12} \sigma_{34}} X_{1+2,3+4} + \frac{(p_1 + p_2 + p_3)^2}{\sigma_{12} \sigma_{23}} \frac{X_{4,5}}{s_{123}} + \frac{(p_2 + p_3 + p_4)^2}{\sigma_{23} \sigma_{34}} \frac{X_{5,1}}{s_{234}}  \\
&= \frac{1}{\sigma_{12} \sigma_{23} \sigma_{34}} \left( \sigma_{23} X_{1+2,3+4} + \sigma_{34} X_{1+2+3,4} + \sigma_{12} X_{1,2+3+4}\right) \\
&=  \frac{p_5^2}{\sigma_{12} \sigma_{23} \sigma_{34}}.
\end{align}
Once again, we see that the amplitude vanishes. 

We will prove that the $X$-amplitudes always vanish by induction.
\begin{figure}[t]
\centering
\includegraphics[width=0.33\textwidth]{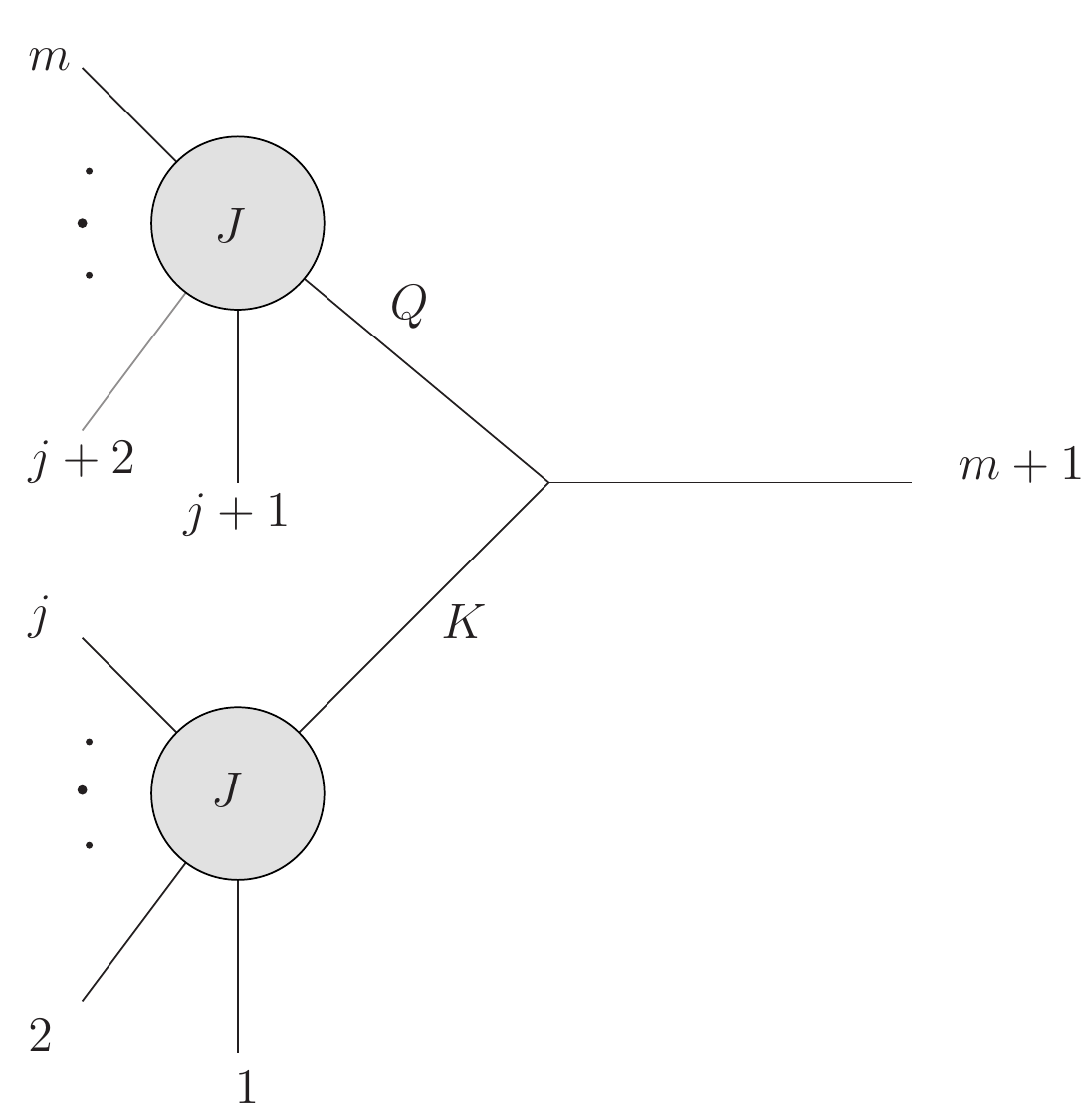}
\caption{Construction of $X$-amplitudes using Berends-Giele recursion.}
\label{fig:bgRecursion}
\end{figure}
These amplitudes can be constructed recursively in terms of graphs with one leg off shell, using Berends-Giele recursion \cite{Berends:1987me}; see Figure~\ref{fig:bgRecursion}. The concept is simple; let us consider a subgraph with $m$ external legs and a single internal leg (from the point of view of the complete graph), which we can call $m+1$. This leg is off-shell until the last stage of the recursion, when we get the complete graph, in which case it is the last remaining external leg. The leg $m+1$ must connect to a three-point vertex. The other legs of this vertex, call them $K$ and $Q$, connect to colour-ordered subdiagrams with a single off-shell leg. Let us define this single off-shell graph, with $j$ on-shell legs and a final leg $K$ off shell to be $J(1, 2, \ldots,j; K)$. By recursion, the object $J$ satisfies
\begin{equation}
J(1, 2, \ldots, m; m + 1) = \sum_{j = 1}^{m - 1} J(1, 2, \ldots, j; K) J(j + 1, j+2, \ldots, m; Q) \frac{1}{K^2} \frac{1}{Q^2} X_{K,Q}.
\end{equation}
The colour-ordered amplitude may simply be obtained as
\begin{equation}
A^{(X)}(1, 2, \ldots, n) = J(1, 2, \ldots, n-1; n) \big|_{p_n^2=0}.
\end{equation}
We adopt the inductive hypothesis that
\begin{equation}
J(1, 2, \dots, m; m + 1) = \frac{1}{\sigma_{12} \sigma_{23} \cdots \sigma_{m-1, m}} \left( \sum_{s = 1}^{m} p_{s}\right)^2
\end{equation}
holds for $m+1\leq n$; this is easily checked for small $n$, as in the four and five-point examples above. Now, consider the $m + 2$ point case, obtained from the recursion formula:
\begin{align}
J(1, 2, \ldots, m+1; m + 2) &= \frac{1}{\sigma_{12} \sigma_{23} \cdots \sigma_{m,m+1}} \sum_{j = 1}^{m} X_{1+\ldots+j,(j+1)+\ldots+(m+1)}\sigma_{j, j+1}, \\
&= \frac{1}{\sigma_{12} \sigma_{23} \cdots \sigma_{m,m+1}} \left( \sum_{s = 1}^{m+1} p_{s}\right)^2,
\end{align}
where we have used the identity Eq.~\eqref{eq:cangemi}. Thus, our inductive hypothesis is proven, and consequently all the $X$-amplitudes (with the exception of the three-point case) vanish identically.

%%%%%%%%%%%%%%%%%%%%%%%%%%%%%%%%%%%%%
%%%%%%%%%%%%%%%%%%%%%%%%%%%%%%%%%%%%%

\section{BCJ numerators from the scattering equations}
\label{sec:numerators}

In this section, we will show how to construct BCJ numerators based on elements of the Lie algebras associated to solutions of the scattering equations. We will start by giving the general idea, and then we will present the complete BCJ numerators. Finally, we will consider a special $SL(2,\Co)$ frame in which the numerators simplify, allowing for a simple proof of their validity.

\subsection{Numerators from commutators: the idea}

The starting point is a very simple observation. Consider three-point scattering, and take vectors of the type
\be
V_a = e^{-ik_a\cdot x} \varepsilon_a \cdot \partial, \qquad a=1,2,3,
\ee
as an example. The most natural Lorentz invariant quantity constructed from the vectors is
\begin{align}
&V_1\cdot[V_2,V_3] + V_2\cdot[V_3,V_1] + V_3\cdot[V_1,V_2] = \nonumber \\
&= -i e^{-i(k_1+k_2+k_3)\cdot x}\,\left( 
(\varepsilon_1 \cdot \varepsilon_2)
(k_1-k_2)\cdot \varepsilon_3 +
(\varepsilon_2 \cdot \varepsilon_3)
(k_2-k_3)\cdot \varepsilon_1 +
(\varepsilon_3 \cdot \varepsilon_1)
(k_3-k_1)\cdot \varepsilon_2
 \right).
\end{align}
Indeed, this is the three-point gluon amplitude. The natural generalisation to four points, in the case of the $s_{12}$-channel, is
\be
\label{eq:eg4ptnumvec}
n_{12,34} = V_1\cdot[V_2,[V_3,V_4]] + V_2\cdot[[V_3,V_4],V_1] + V_3\cdot[V_4,[V_1,V_2]] + V_4\cdot[[V_1,V_2],V_3],
\ee
which corresponds to the colour factor $\tf^{a_1a_2b}\tf^{ba_3a_4}$. Notice how the structure of the commutators reflects the orientation of the vertices. Viewing this quantity as a numerator, it turns out that the BCJ Jacobi identities follow from the standard Jacobi identities of the algebra of spacetime vectors,
\begin{align}
&n_{12,34}+n_{23,14}+n_{31,24} = \nonumber \\
&= V_1\cdot \left( [V_2,[V_3,V_4]] + [V_3,[V_4,V_2]] + [V_4,[V_2,V_3]] \right) \nonumber \\
&+ V_2\cdot \left( [V_3,[V_1,V_4]] + [V_1,[V_4,V_3]] + [V_4,[V_3,V_1]] \right) \nonumber \\
&+ V_3\cdot \left( [V_1,[V_2,V_4]] + [V_2,[V_4,V_1]] + [V_4,[V_1,V_2]] \right) \nonumber \\
&+ V_4\cdot \left( [[V_1,V_2],V_3] + [[V_2,V_3],V_1] + [[V_3,V_1],V_2] \right) =0.
\label{eq:jacobivvvv}
\end{align}
It is straightforward to generalise this construction for any number of external particles, so that the Jacobi identities hold. Take the example \eqref{eq:eg4ptnumvec}. If we substitute $V_r$ by $T^{a_r}$ and the $\cdot$ product by standard matrix multiplication, and then take the trace, we find that each term reproduces the corresponding colour factor. This will be the general rule. So, the numerator of the trivalent graph $\alpha$ is
\be
\label{numcom}
n_\alpha = \sum_{a=1}^n V_a \cdot \mathfrak{G}_a^{(\alpha)},
\ee
where $\mathfrak{G}_a^{(\alpha)}$ is the commutator structure of graph $\alpha$ as read from particle $a$. Let us give another example, the seven-point graph with colour factor
\be
\tf^{a_1 a_2 b}\tf^{b a_3 c}\tf^{c d e}\tf^{d a_4 a_5}\tf^{e a_6 a_7}.
\ee
Its numerator is 
\begin{align}
n &= V_1\cdot[V_2,[V_3,[[V_4,V_5],[V_6,V_7]]]] +  V_2\cdot[[V_3,[[V_4,V_5],[V_6,V_7]]],V_1] \nonumber \\
&+ V_3\cdot[[[V_4,V_5],[V_6,V_7]],[V_1,V_2]] \nonumber \\
&+ V_4\cdot[V_5,[[V_6,V_7],[[V_1,V_2],V_3]]] + V_5\cdot[[[V_6,V_7],[[V_1,V_2],V_3]],V_4] \nonumber \\
&+ V_6\cdot[V_7,[[[V_1,V_2],V_3],[V_6,V_7]]] + V_7\cdot[[[[V_1,V_2],V_3],[V_6,V_7]],V_6].
\end{align}

Our BCJ numerators for gluon amplitudes will be based on the same idea, although the algebras do not admit this representation in general. The idea described here follows from the self-dual story that we saw before, and it was presented in \cite{talks}, where it was claimed to apply directly to MHV amplitudes, which is a particular case of the analysis to be presented next. The idea was also presented independently in \cite{Fu:2012uy}, where it was used to obtain basis amplitudes in the spirit of \cite{BjerrumBohr:2012mg}.

\subsection{BCJ numerators from kinematic algebras}

We will construct BCJ numerators for each contribution to the gauge theory amplitude corresponding to a solution of the scattering equations. According to \eqref{eq:amplitudegauge}, each of these contributions is of the type
\be
\label{eq:PTperm}
\text{Parke-Taylor factor} \quad \times \quad \text{permutation-invariant factor}.
\ee
The permutation-invariant factor can be pulled out. Therefore, we are really looking for BCJ numerators reproducing the Parke-Taylor amplitudes,
\be
\label{eq:APT}
{\mathcal A}_{\textrm{PT}}^{(I)} = \frac{\text{Tr}(T^{a_1}T^{a_2}\cdots T^{a_n})}{\sigma_{12}^{(I)}\cdots\sigma_{n1}^{(I)}} + \text{non-cyclic permutations},
\ee
where $I$ denotes the particular solution to the scattering equations. Below, we will omit the label $I$ in order to simplify the notation.

The Lie algebras \eqref{eq:hvLie} associated to each solution of the scattering equations are not of the same type as the Lie algebra discussed in the subsection above. However, the latter will serve as inspiration. First, it will be convenient to define the action of an element of the Lie algebra on another as, in the case of the $X$-algebra,
\be
\label{eq:rulecom1}
\hv^+_A \hv^+_B = i X_{A,B} \hv^+_{A|B}.
\ee
Here, $\hv^+_{A|B}$ is not itself an element of the $X$-algebra, but satisfies the property
\be
\hv^+_{A|B} + \hv^+_{B|A} = \hv^+_{A+B}.
\ee
This is consistent with $[\hv^+_A,\hv^+_B] = i X_{A,B} \hv^+_{A+B}$, and it clearly mirrors the vector algebra seen above, where \eqref{eq:rulecom1} corresponds to the first part of the vector commutator. Based on this idea, and further specifying $\hv^\pm_{\emptyset|A}=\hv^\pm_A$, we define
\begin{align}
\label{actionV}
\hv^+_{A|B} \hv^\pm_{C|D} = i X_{B,C+D} \hv^\pm_{A+B+C|D}, \\
\hv^-_{A|B} \hv^\pm_{C|D} = i \Xb_{B,C+D} \hv^\pm_{A+B+C|D}.
\end{align}
Moreover, we define a symmetric product $\ast$ such that
\be
\hv^\pm_{A|B} \ast \hv^\pm_{C|D} = 0, \qquad \,\hv^+_{A|B} \ast \hv^-_{C|D} = \left\{ 
\begin{array}{rl} - 2h_B h_D & \text{     if } K_A+K_B+K_C+K_D=0, \\ 0\phantom{aaaaa} & \text{      otherwise.} \end{array} \right.
\ee
where $h_A =\sum_{a\in \{A\}} h_a$.

The sole goal of these definitions, for our purposes, is to determine objects such as, at four points,
\be
\hv^-_1 \ast  \hv^+_2 \hv^+_3 \hv^+_4 = 2 X_{2,3+4} X_{3,4} h_1h_4=2 X_{1,2}X_{3,4} h_1h_4,
\ee
where we used the scattering equations. If we consider commutators, and recall the conditions \eqref{eq:condh} on the $h_a$, we get
\be
\hv^-_1 \ast  [\hv^+_2,[\hv^+_3,\hv^+_4]] =2 X_{1,2} X_{3,4} h_1(h_2+h_3+h_4)=-2 h_1^2 X_{1,2} X_{3,4}.
\ee
We can now consider complete numerators such as
\be
n_{1^-2^+,3^+4^+} =\hv^-_1 \ast  [\hv^+_2,[\hv^+_3,\hv^+_4]] + \hv^+_2 \ast  [[\hv^+_3,\hv^+_4],\hv^-_1] +\hv^+_3 \ast  [\hv^+_4,[\hv^-_1,\hv^+_2]] + \hv^+_4 \ast  [[\hv^-_1,\hv^+_2],\hv^+_3],
\ee
which satisfy Jacobi identities exactly for the same reason as in \eqref{eq:jacobivvvv}. The extension of this construction of numerators to $n$ points is straightforward, as pointed out in the previous subsection.

Regarding numerators where only one of the $\hv$'s is of the $(-)$ type, as in the example above, there are two important observations to make, which extend to any number of particles.\footnote{Notice that $(\pm)$ does not refer to the polarisation of the external particles. That information is included in the permutation-invariant factor, as mentioned in \eqref{eq:PTperm}. The labels $(\pm)$ will be used only in reproducing the Parke-Taylor factor.}
The first is that we do not need to consider a sum over all the particles when we take $\hv_a \ast \text{[commutators]}$. For instance, in our example,
\be
n_{1^-2^+,3^+4^+} = - 4h_1^2 X_{1,2} X_{3,4} =2\, \hv^-_1 \ast  [\hv^+_2,[\hv^+_3,\hv^+_4]] .
\ee
In general, the sum over all the particles is equivalent to considering only the $(-)$ particle, up to a factor of two; i.e. for graph $\alpha$, if $r$ labels the $(-)$ particle, we have
\be
\label{eq:numrsamem}
n_{\alpha,r} = \sum_{a=1}^n \hv_a \ast \hat{\mathfrak{G}}_a^{(\alpha)} = 2 \hv^-_r \ast \hat{\mathfrak{G}}_r^{(\alpha)}.
\ee
We have verified this numerically, but it would be nice to have a proof. 

The second observation is that numerators with a single $(-)$ particle contain only $X$ vertices, apart from an overall factor proportional to $h_r^2$. Therefore, these are the numerators of what we called $X$-amplitudes. We have proven previously that these amplitudes vanish, even though the numerators do not. Notice also that, if we only had $(+)$ particles, the numerators themselves would vanish due to the product $\ast$.

We now proceed to the main result of this paper. The Parke-Taylor amplitude \eqref{eq:APT} is obtained with a very simple prescription: we consider numerators of the type described above but with two $(-)$ particles. Let these particles be $r$ and $s$. We obtain
\be
\label{eq:APTnum}
{\mathcal A}_{\textrm{PT}} = \alpha_{rs} \sum_{\alpha\in \textrm{cubic}} \frac{n_{\alpha,\,rs} c_\alpha}{D_\alpha},
\ee
where the factor $\alpha_{rs}$, which is independent of the particle ordering, is given by
\be
\alpha_{rs} =-\left[ 4 i^n(\sigma_r-\sigma_s)^2 h_r h_s (h_r+h_s) \sum_{a=1}^n \sigma_a^2 h_a \right]^{-1}.
\ee
The special dependence of $\alpha_{rs}$ on particles $r$ and $s$ compensates that of the numerators $n_{\alpha,\,rs}$, so that the choice of these particles is irrelevant for ${\mathcal A}_{\textrm{PT}}$. We view this fact as an analogue of choosing arbitrarily two columns/rows to be eliminated in the $\text{Pf}\;'\Psi$ defined in \eqref{eq:pfPsi}. For two $(-)$ particles, we find the property analogous to \eqref{eq:numrsamem} that
\be
n_{\alpha,rs} = \sum_{a=1}^n \hv_a \ast \hat{\mathfrak{G}}_a^{(\alpha)} = 
2 \left( \hv^-_r \ast \hat{\mathfrak{G}}_r^{(\alpha)} + \hv^-_s \ast \hat{\mathfrak{G}}_s^{(\alpha)} \right).
\ee
We emphasise that these results do not depend on the values of the quantities $h_a$, as long as they obey the conditions \eqref{eq:condh}.

We conclude that a natural choice of BCJ numerator for graph $\alpha$ in a gauge theory amplitude is
\be
\label{eq:bcjnumfinal}
n_\alpha = \sum_{I=1}^{(n-3)!} \alpha_{rs}^{(I)} n_{\alpha,\,rs}^{(I)} \Upsilon^{(I)}, \qquad \text{with} \quad
\Upsilon^{(I)}=\frac{\text{Pf}\;'\Psi^{(I)}}{\text{det}\;'\Phi^{(I)}},
\ee
where we reintroduced the label $I$ of each solution to the scattering equations.
Since the factors $\alpha_{rs}^{(I)}$ and $\Upsilon^{(I)}$ are independent of the particle ordering, the complete BCJ numerators $n_\alpha$ satisfy the same Jacobi identities as the numerators $n_{\alpha,\,rs}^{(I)}$.

\subsection{Proof of BCJ numerators with reference particle}

We have presented above the complete BCJ numerators based on the kinematic algebras. Their validity has been verified numerically up to eight points. In the following, we will use a certain $SL(2,\Co)$ frame in order to obtain a simpler form of the numerators, which allows us to prove their validity for any multiplicity. This form of the numerators reduces to the one described in \cite{Monteiro:2011pc} for MHV amplitudes.

We mentioned above that the numerators obtained with the elements $\hv^\pm_a$ are, in the case of one $(-)$ particle, the numerators of an $X$-amplitude (which vanishes), and, in the case of two $(-)$ particles, the numerators of a Parke-Taylor amplitude (up to a permutation-invariant factor). In the case of one $(-)$ particle, there are only $X$ vertices, while in the case of two $(-)$ particles there is a single $\Xb$ vertex, the rest being $X$ vertices. This is easily checked by direct inspection. We are interested in the case of two $(-)$ particles, and we want to choose an $SL(2,\Co)$ frame such that we force the single vertex $\Xb$ to be attached to a certain external particle. Let us say that this reference particle is particle $n$. This is achieved by taking the limit
\be
\sigma_n \to \infty.
\ee
In this limit, due to the second condition in \eqref{eq:condh}, we must have also $h_n\to0$, so that $\sigma_n h_n$ is finite. Therefore, we have
\be
X_{n,A} \to 0, \qquad \Xb_{n,A} \to (\sigma_n h_n) h_A,
\ee
which leads to the vanishing of all the contributions for which particle $n$ is attached to an $X$ vertex, rather than the $\Xb$ vertex. It is then trivial to write down the BCJ numerators: they are the same as for the $X$-amplitudes, except that one of the vertices -- the one attached to particle $n$ -- is $\Xb$. Therefore, all the Jacobi identities involving only $X$ vertices are satisfied. The only additional requirement is that the Jacobi identities involving propagators connected to particle $n$ are also satisfied. If the other three (generically off-shell) lines connected to such a propagator are $A$, $B$ and $C$, we have
\begin{align}
&\Xb_{n,A} X_{B,C} + \Xb_{n,B} X_{C,A} +  \Xb_{n,C} X_{A,B} \nonumber \\
&\qquad \to (\sigma_n h_n) \left( h_A X_{B,C} + h_B X_{C,A} + h_C X_{A,B} \right) \nonumber \\
&\qquad = (\sigma_n h_n) \left( h_A (X_{B,C}+X_{B,A}) + h_B (X_{C,A}+X_{B,A}) - h_n X_{A,B} \right) \nonumber \\
&\qquad = (\sigma_n h_n) \left( h_A X_{n,B} + h_B X_{A,n} - h_n X_{A,B} \right) \nonumber \\
&\qquad \to 0,
\end{align}
where we used the condition $\sum_a h_a=0$ in the third line and the scattering equations in the fourth line.

We now proceed to prove the validity of these BCJ numerators using Berends-Giele recursion. The procedure is essentially the same as in Section~\ref{subsec:proofXvanish}, except that the final vertex in the recursion is not $X$, but $\Xb$. Therefore,
\begin{align}
A_{\text{PT}}(1, 2, \ldots, n) &= \alpha_n \sum_{j = 1}^{n - 2} J(1, 2, \ldots, j; K) J(j + 1, j+2, \ldots, n-1; Q) \frac{1}{K^2} \frac{1}{Q^2} \bar X_{n,K} \\
&= \alpha_n \frac{1}{\sigma_{12} \sigma_{23} \cdots \sigma_{n - 2,n-1}}  \sum_{j = 1}^{n - 2} \bar X_{n, 1+\ldots j} \sigma_{j, j+1},
\end{align}
where $\alpha_n$ is a proportionality coefficient independent of particle ordering.
It is useful to rearrange the summation on the last line:
\begin{equation}
\sum_{j = 1}^{n - 2} \bar X_{n, 1+\ldots j} \sigma_{j, j+1} =
 \sum_{1 \leq s \leq j < n-1} \bar X_{n,s} \sigma_{j, j+1} 
= \sum_{s = 1}^{n - 1}  \bar X_{n,s} (\sigma_s - \sigma_{n-1}) 
= \sum_{s = 1}^{n - 1}  \bar X_{n,s} \sigma_s,
\end{equation}
where we used the equation \eqref{SEXXbar} for $\Xb$ in the last equality. In our limit, we can use $\sum_a \sigma_a h_a=0$ to see that
\begin{equation}
\sum_{j = 1}^{n - 2} \bar X_{n, 1+\ldots j} \sigma_{j, j+1} = \sigma_n h_n \sum_{j = 1}^{n - 1} h_s\sigma_s 
= - (\sigma_n h_n)^2,
\end{equation}
which is a finite quantity. Finally, if we set
\begin{equation}
\alpha_n = \frac{1}{\sigma_n^2(\sigma_n h_n)^2},
\end{equation}
we obtain the correct Parke-Taylor amplitude in our limit,
\begin{equation}
A_{\text{PT}}(1, 2, \ldots, n) = \frac{1}{\sigma_{12} \sigma_{23} \cdots \sigma_{n - 2,n-1}(-\sigma_n)\sigma_n}.
\end{equation}

%%%%%%%%%%%%%%%%%%%%%%%%%%%%%%%%%%%%%
%%%%%%%%%%%%%%%%%%%%%%%%%%%%%%%%%%%%%

\section{Gravity amplitudes and colour-dual traces}
\label{sec:gravity}

In the previous section, we have obtained BCJ numerators for gauge theory amplitudes. The BCJ double-copy to gravity guarantees that we obtain also expressions for gravity amplitudes. In particular, we have, from the result \eqref{eq:bcjnumfinal},
\begin{equation}
(-2)^{n-3}\mathcal{M}_n = \sum_\alpha \frac{n_\alpha n_\alpha}{D_\alpha} = \sum_{I=1}^{(n-3)!} \sum_{J=1}^{(n-3)!} 
\alpha_{rs}^{(I)}\alpha_{rs}^{(J)} \Upsilon^{(I)} \Upsilon^{(J)}
\sum_\alpha \frac{n_{\alpha,\,rs}^{(I)} n_{\alpha,\,rs}^{(J)}}{D_\alpha},
\qquad \Upsilon^{(I)}=\frac{\text{Pf}\;'\Psi^{(I)}}{\text{det}\;'\Phi^{(I)}}.
\end{equation}
The BCJ counterpart of KLT orthogonality is
\begin{equation}
\label{eq:bcjortho}
\sum_\alpha \frac{n_{\alpha,\,rs}^{(I)} n_{\alpha,\,rs}^{(J)}}{D_\alpha} =0 \quad \text{for} \quad I\neq J.
\end{equation}
This leads to the simplification
\begin{equation}
(-2)^{n-3}\mathcal{M}_n %= \sum_{I=1}^{(n-3)!}
%({\alpha_{rs}^{(I)}} {\Upsilon^{(I)}})^2
%\sum_\alpha \frac{n_{\alpha,\,rs}^{(I)} n_{\alpha,\,rs}^{(I)}}{D_\alpha}
=\sum_{I=1}^{(n-3)!} \left( \frac{\text{Pf} \, ' \Psi^{(I)}}{\text{det}\;'\Phi^{(I)}} \right)^2
\;{\alpha_{rs}^{(I)}}^2
\sum_\alpha \frac{n_{\alpha,\,rs}^{(I)} n_{\alpha,\,rs}^{(I)}}{D_\alpha}.
\end{equation}
Agreement with the expression \eqref{eq:amplitudegrav} implies that
\begin{equation}
{\alpha_{rs}^{(I)}}^2 \sum_\alpha \frac{n_{\alpha,\,rs}^{(I)} n_{\alpha,\,rs}^{(I)}}{D_\alpha} = (-2)^{n-3}\text{det}\;' \Phi^{(I)}.
\end{equation}
We have verified this result numerically up to eight points. Therefore, we have obtained a natural expansion for the Jacobian $\text{det}\;' \Phi^{(I)}$ in terms of trivalent graphs.

The kinematic algebras allow us to perform a different type of squaring to gravity, proposed in \cite{Bern:2011ia}. This is based on the substitution of the standard colour traces of gauge theory, $\text{Tr}(T^{a_1}T^{a_2}\cdots T^{a_n})$, by kinematic analogues, $\tau_{(1,2,\ldots,n)}$, which share the cyclic symmetry. The gauge theory amplitude can be written in terms of the ``kinematic traces", or colour-dual traces \cite{Bern:2011ia}, as
\begin{align}
{\mathcal A}_n &=\sum_{\text{non-cyclic}}  \textrm{Tr}(T^{a_1}T^{a_2}\cdots T^{a_n}) \, A(1,2,\ldots,n) \nonumber \\
&= \sum_{\text{non-cyclic}}  \tau_{(1,2,\ldots,n)} \, \theta(1,2,\ldots,n),
\label{eq:dualdecomp}
\end{align}
where the sums run over non-cyclic permutations of the external labels. Here, $\theta$ is a dual amplitude constructed from the colour-ordered gauge theory amplitude $A$ by replacing all kinematic numerators with color factors, {\it i.e.} $n_i \to c_i$. We can see $\theta$ as the amplitude of a cubic massless scalar theory with two Lie groups that is colour-ordered with respect to one of the groups; we mentioned these theories in Sections~\ref{sec:SE} and \ref{sec:review}. For these theories, we actually have the double-colour-dressed amplitude
\be
\label{gravtau}
\Theta_n =\sum_{\text{non-cyclic}}  \textrm{Tr}(T^{a_1}T^{a_2}\cdots T^{a_n}) \, \theta(1,2,\ldots,n).
\ee
The fact that we can write a gauge theory amplitude in the dual form \eqref{eq:dualdecomp} implies that the gravity amplitude can be written as \cite{Bern:2011ia}
\be
\label{gravtau}
(-2)^{n-3}{\mathcal M}_n =\sum_{\text{non-cyclic}}  \tau_{(1,2,\ldots,n)} \, A(1,2,\ldots,n).
\ee
Constructions for valid $\tau_{(1,2,\ldots,n)}$'s were also presented in \cite{BjerrumBohr:2012mg,Du:2013sha,Fu:2013qna}.

From our discussion in the previous section, it is easy to write down canonical expressions for these ``kinematic traces". We use objects constructed from elements of the $(+)$ and $(-)$ Lie algebras. Let us define, for a given solution to the scattering equations,
\begin{equation}
\tau_{(1,2,\ldots,n),rs} =\hv_1 \ast \hv_2\hv_3\ldots\hv_n  + \text{cyclic permutations},
\end{equation}
where two of the particles are of the $(-)$ type, specifically particles $r$ and $s$, while the remaining are $(+)$ particles. We find (numerically up to eight points) that
\begin{equation}
\alpha_{rs} \sum_{\text{cyclic}} \frac{\tau_{(1,2,\ldots,n),rs}}{\sigma_{12}\cdots\sigma_{n1}} = (-2)^{n-3}\text{det}\;' \Phi.
\end{equation}
Finally, reintroducing the label $I$ for each solution to the scattering equations, we can write the complete ``kinematic traces" as
\be
\tau_{(1,2,\ldots,n)} = \sum_{I=1}^{(n-3)!} \alpha_{rs}^{(I)} \tau_{(1,2,\ldots,n),rs}^{(I)} \Upsilon^{(I)}.
\ee
We can also point out that the counterpart of KLT orthogonality in terms of ``kinematic traces" is
\begin{equation}
\label{eq:traceortho}
\sum_{\text{cyclic}} \frac{\tau_{(1,2,\ldots,n),rs}^{(I)}}{\sigma_{12}^{(J)}\cdots\sigma_{n1}^{(J)}} = 0
 \quad \text{for} \quad I\neq J.
\end{equation}

For completeness, we also comment on the double-copy of the BCJ numerators corresponding to the $X$-amplitudes. Just like in the well-known four-dimensional self-dual case, although the numerators constructed only from $X$ vertices don't vanish, the double-copy vanishes. If we consider the expressions \eqref{eq:bcjortho} or \eqref{eq:traceortho}, but take the numerators and the ``kinematic traces" to be constructed with a single $(-)$ particle, rather than two, then the vanishing holds for any $I$ and $J$.

%%%%%%%%%%%%%%%%%%%%%%%%%%%%%%%%%%%%%
%%%%%%%%%%%%%%%%%%%%%%%%%%%%%%%%%%%%%

\section{Discussion}
\label{sec:disc}

We have shown how the scattering equations can be interpreted as a consistency condition for a quantity which plays the role of a vertex,
$$
X_{a,b} = \frac{s_{ab}}{\sigma_a-\sigma_b}.
$$
We explored the close relationship of this quantity with four-dimensional self-dual gauge theory, whose vertex is a particular case of the general story presented here. A crucial characteristic of the self-dual vertex -- the fact that it corresponds to the structure constant of a Lie algebra -- was seen to extended to the general case. Moreover, we have demonstrated that the $X$-amplitudes, which we defined as ``amplitudes" constructed solely from the $X$ vertex and scalar propagators, always vanish, just like tree-level amplitudes in the self-dual theory.

Using the $X$ vertex and a companion $\Xb$ vertex, which we defined appropriately, we were able to put together the elements of the respective Lie algebras in order to construct BCJ numerators for the scattering amplitudes of gauge theory. An interesting problem would be to obtain explicit representations for the elements $\hv^{\pm}_A$ of these Lie algebras which reproduce our prescriptions. Nevertheless, the prescriptions have also allowed us to define the kinematic counterparts of colour traces. Finally, we saw how the double-copy to gravity, performed either with trivalent graphs or with ``kinematic traces", provides natural decompositions of the determinant $\text{det}\;'\Phi$, which arises from the integration measure of the scattering amplitudes.

The results presented here offer a new insight into the extension to loop level of the formulas for amplitudes based on the scattering equations. We point out that an object defined essentially as $X_{a,b}$, which satisfies both the Jacobi identities and the scattering equations, plays a role analogous to a vertex in one-loop superstring amplitudes \cite{Mafra:2012kh}. In the case of self-dual gauge theory (and self-dual gravity), it is already known that the kinematic algebras allow for the construction of one-loop rational amplitudes \cite{Boels:2013bi}; and there is a closely related story for one-loop amplitudes in ${\mathcal N}=4$ super-Yang-Mills theory \cite{Bjerrum-Bohr:2013iza}.

\appendix

\acknowledgments
We thank Lionel Mason, Gustav Mogull, V. Parameswaran Nair and David Skinner for helpful discussions. RM is supported by the European Commission through a Marie Curie Fellowship, while DOC is supported in part by the STFC grant Particle Physics at the Tait Institute.

\bibliographystyle{jhep}

\end{document}